\documentclass[aps,prc,groupedaddress,showpacs,twocolumn,floatfix]{revtex4}

\usepackage{graphicx}

\def\gtorder{\mathrel{\raise.3ex\hbox{$>$}\mkern-14mu
 \lower0.6ex\hbox{$\sim$}}}
\def\ltorder{\mathrel{\raise.3ex\hbox{$<$}\mkern-14mu
 \lower0.6ex\hbox{$\sim$}}}
\def\mugegm{\mu_p G_E / G_M}

\def\ge{G_E}
\def\gm{G_M}
\def\se{\sigma_E}
\def\sm{\sigma_M}

\def\beq{\begin{equation}}
\def\eeq{\end{equation}}
\def\ba{\begin{eqnarray*}}
\def\ea{\end{eqnarray*}}

\begin{document}

\title{Coulomb distortion in high-$Q^2$ elastic $e$--$p$ scattering}

\author{John Arrington}
\affiliation{Physics Division, Argonne National Laboratory, Argonne, Illinois 
60439, USA}

\author{Ingo Sick}
\affiliation{Dept. f\"{u}r Physik und Astronomie, Universit\"{a}t Basel, CH-4056
Basel, Switzerland}

\date{\today}

\begin{abstract}

Recently, there has been a significant amount of activity to try and understand
the discrepancy between Rosenbluth and polarization transfer measurements of
the proton form factors.  It has been suggested that the standard use of
plane-wave Born approximation in extracting the form factors is insufficient,
and that higher-order terms must also be included. Of the corrections missing
in standard prescriptions, Coulomb distortion is the most well understood. In
this paper, we examine the effect of Coulomb distortion on the extraction of
the proton form factors.

\end{abstract}

\pacs{25.30.Bf, 13.40.Gp}

\maketitle

\section{Introduction}
The determination of the proton charge form factor $\ge$ recently has received
extensive attention. The determination via the traditional 'Rosenbluth'
separation of the longitudinal (L, charge) and transverse (T, magnetic)
contributions to the electron-proton elastic cross section disagrees at large
momentum transfer $q$ with the results obtained via measurements of the proton
recoil polarization \cite{jones00}. It was immediately suspected that this
could come from the fact that at large $q$ the magnetic cross section $\sm$
becomes much larger than the electric cross section $\se$. Therefore, small
corrections to the cross section yield significant corrections to the small
contribution proportional to $\ge^2$.

A reanalysis of the available data on $e$--$p$ scattering with careful
consideration of the systematic errors has confirmed the presence of the
discrepancy \cite{arrington03a}. A recent experiment \cite{arrington04c}
exploiting a novel technique less sensitive to systematic errors --- recoil
detection in the L/T-separation --- has also confirmed the discrepancy between
the values of $\ge$ from L/T-separation and recoil-polarization measurement

Traditionally, $\ge$ and $\gm$ have been determined using the Rosenbluth
technique and the plane-wave Born approximation, PWIA. Several papers have
recently pointed out that second-order effects could play an important role
\cite{blunden03,guichon03,chen04}. Among the second order effects two
particular ones can be singled out:

1. The effect of the proton charge to Coulomb distortion of the ingoing and
outgoing electron waves. This distortion has traditionally been included in the
analysis of electron scattering experiments for nuclei with $Z$$>$1, but almost
always neglected for $Z$=1 where it has been effectively absorbed into the form
factors. For $Z$=1, Coulomb distortion has been shown to have significant
effects on the proton \textit{rms}-radius \cite{sick03, rosenfelder00} and to
remove a longstanding discrepancy for the deuteron \cite{sick98}.
Diagrammatically, Coulomb distortion corresponds to the exchange of one hard
and one (or several) soft photons.
 
2. The exchange of two hard photons.  This contribution plays a role mainly at
the larger $q$'s. Because $\sm \gg \se$, one could expect that the dominant
term comes from two successive magnetic (\textit{i.e.} spinflip) interactions
which therefore contribute to the small electric (\textit{i.e.} non-spinflip)
term. Indeed, some exploratory calculations \cite{blunden03,tjon_priv} support
this scenario. Tjon has calculated the second order contribution with the
proton magnetic moment, $\mu_p$, reduced from 2.79 to 1. This modification
greatly reduces the second-order effect, the left-over contribution being of
the order of what is expected from Coulomb-distortion alone. This
qualitatively can be understood in terms of the above model; if two successive
magnetic scatterings were the only process, the contribution would be expected
to scale with
$\mu_p^2$.

Coulomb distortion is an established effect and fairly straightforward to
calculate \cite{sick98,rosenfelder00}. More difficulties arise for the case of
the exchange of two hard photons. Here not only the proton ground state
(treated in \cite{blunden03}), but also the proton excited states come in.
These dispersive effects are much more difficult to calculate. In this paper
we study the effect of Coulomb distortion \textit{alone}. Among the various
second-order contributions, this is the one which is best established.

\section{Coulomb distortion}

We have calculated the Coulomb distortion correction for electron-proton
scattering using the second Born approximation, following the approach
presented in Ref.~\cite{sick98} for electron-deuteron scattering. This series
in $Z\alpha$ is expected to be very accurate for $Z\alpha \sim 0.01$ of
interest for hydrogen.

The corrections have been calculated using an exponential charge distribution
for the proton, in accordance with the fact that the proton charge form factor
is close to a dipole. The deviations from the dipole shape found at very high
$q$ are not expected to have consequences on the Coulomb distortion, which is
a long-range effect. One finds that the effect of the Coulomb distortion is of
order 1\% of the cross section. It is mainly dependent on angle, and thus does
have effects in Rosenbluth separations which compare data at the same $q$ but
different angles.


\section{Results}

\begin{figure}[htb]
\includegraphics[height=5.5cm,width=8.0cm,angle=0]{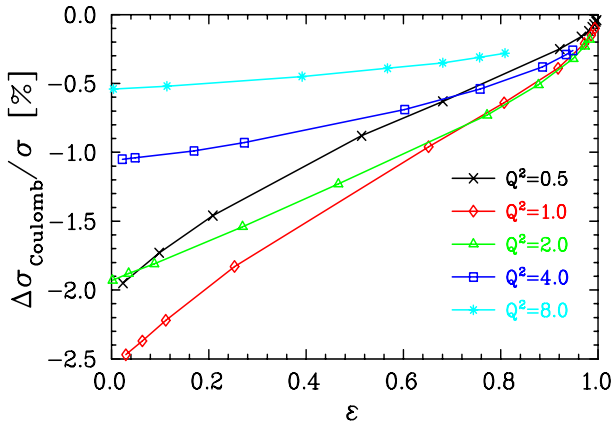}
\caption{(Color online) Coulomb distortion to the elastic electron-proton
cross section.
\label{fig:dsigma}}
\end{figure}

Figure~\ref{fig:dsigma} shows the effect of the Coulomb distortion on
the cross section as a function of $\varepsilon$ for several different
$Q^2$ values.  The effect is maximum near $Q^2 = 1$~GeV$^2$, and has a
significant $\varepsilon$-dependence. Note that the effect is very
nearly linear for these $Q^2$ values, except for the very largest
$\varepsilon$ values.

\begin{figure}[htb]
\includegraphics[height=5.5cm,width=8.0cm,angle=0]{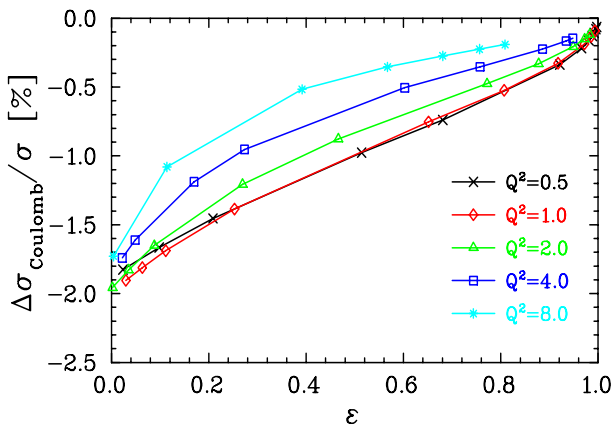}
\caption{(Color online) Coulomb distortion to the elastic electron-proton
cross section in the EMA.
\label{fig:dsigma_ema}}
\end{figure}

Figure~\ref{fig:dsigma_ema} shows the results obtained in an effective
momentum approximation (EMA) calculation.  The EMA yields similar results for
large $\varepsilon$ and large $Q^2$, but is significantly different from the
second-order Born calculation elsewhere. The correction is calculated by
increasing the energy of the incoming and scattered electrons at the interaction
vertex by the Coulomb potential, and evaluating the cross section using
using form factors at the modified $Q^2$ value, but leaving the Mott
cross section unchanged.  For nuclei, the Coulomb potential is usually
determined by assuming a uniform sphere with an \textit{rms}-radius that
matches electron scattering measurements, and then using the potential at the
surface (or center) of the sphere.  For the proton, the uniform sphere 
will tend to underestimate the effect since the charge is more highly
concentrated in the center, so we use 1.9 MeV, the potential at the center.

Alternative prescriptions for the EMA use the modified kinematics for the full
cross section, rather than just the form factors, or else include a focusing
factor. However, in both cases one obtains a significant reduction in the
size of the correction at low $\varepsilon$, improving the agreement somewhat
for very large $Q^2$ values, but making it significantly worse for lower $Q^2$
values. One can also use a different value for the value of the Coulomb
potential to increase or decrease in the correction, but the overall agreement
would not be any better. While the EMA prescription can be `tuned' to give
good agreement for a range in $\varepsilon$ or $Q^2$, none of these approaches
yield an adequate approximation to the exact calculation.

\begin{figure}[htb]
\includegraphics[height=5.5cm,width=8.0cm,angle=0]{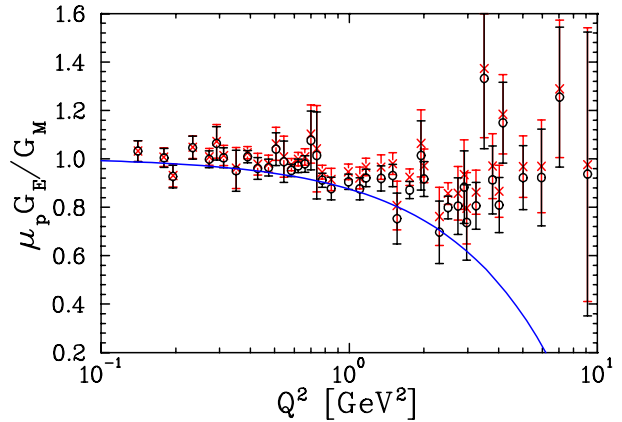}
\caption{(Color online) Rosenbluth extraction of $\mugegm$ before (`x') and
after (circle) correcting for Coulomb distortion.  The solid line shows the
parameterization of the polarization transfer measurements.
\label{fig:gegm}}
\end{figure}

\begin{figure}[htb]
\includegraphics[height=4.5cm,width=8.0cm,angle=0]{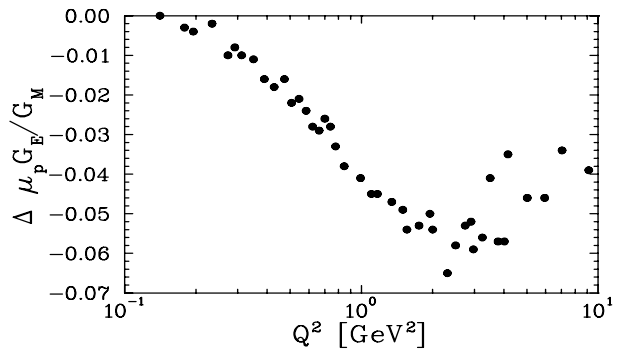}
\caption{The change in the extracted value of $\mugegm$ resulting from
correcting for the effects of Coulomb distortion.
\label{fig:dgegm}}
\end{figure}

The effect of the Coulomb distortion on the form factors at very low $Q^2$
values has been studied in detail~\cite{sick03}.  For larger $Q^2$ values, the
distortion grows with $Q^2$ until approximately 1~GeV$^2$, and then begin to
decrease. However, as $Q^2$ increases, the electric form factor yields a
decreasing $\varepsilon$-dependence in the reduced cross section, and so the
effect of the Coulomb distortion on $\ge$ is magnified as $Q^2$ increases.
Figure~\ref{fig:gegm} shows the Rosenbluth extraction of $\mugegm$ from the
global analysis of Ref.~\cite{arrington04a} before and after correcting the
cross section data for Coulomb distortion.  Applying this correction reduces
the extracted form factor ratio, improving the agreement with polarization
transfer results, but the ratio is still well above the polarization transfer
result. Figure~\ref{fig:dgegm} shows the change in the ratio $\mugegm$ as a
function of $Q^2$.  Because the correction is not precisely linear in
$\varepsilon$, the effect on $\ge$ depends somewhat on the $\varepsilon$ range
covered at each $Q^2$ value, and thus the correction shows a significant
amount of scatter. For intermediate $Q^2$ values, the change in the extracted
value of $\mugegm$ can be as large or larger than the experimental
uncertainties.

\begin{figure}[htb]
\includegraphics[height=5.5cm,width=8.0cm,angle=0]{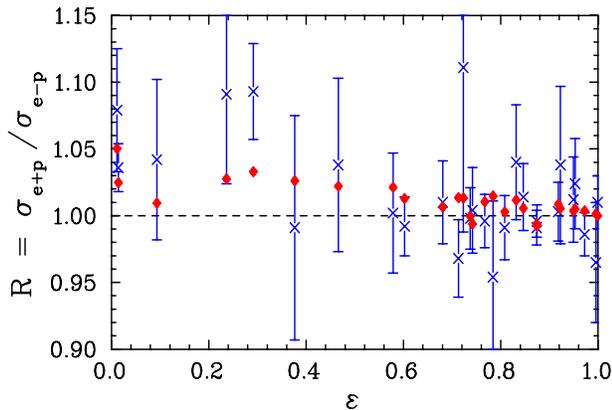}
\caption{(Color online) Measurements of the ratio of positron-proton
to electron-proton scattering cross section (blue `x'), along with the
ratio predicted by the effects of Coulomb distortion (red diamond).
\label{fig:positron}}
\end{figure}

Finally, because the sign of the Coulomb distortion depends on the product of
the beam and target charge, a comparison of electron-proton and positron-proton
scattering is sensitive to these corrections.  Figure~\ref{fig:positron} shows
the ratio of positron-proton cross section to electron-proton cross section
for a series of comparisons made in the 1960s (see Ref.~\cite{arrington04b}
and references therein).  While the uncertainties are large and the results
consistent with $R=1$ ($\chi^2=23.9$ for 28 data points), the data is in
better agreement with the values expected when including Coulomb distortion
($\chi^2=14.7$). In particular, Coulomb distortion reproduces the systematic
enhancement of the ratio at low $Q^2$ and $\varepsilon$.

\section{Conclusions}

In this paper, we have shown that Coulomb distortion has a non-negligible
effect on the proton elastic cross section.  The main effect is a change in
the $\varepsilon$-dependence of the cross section. The $\varepsilon$-dependent
correction behaves approximately as $1/Q^2$ for $Q^2>2$~GeV$^2$, as does
the contribution from $\ge$.  Thus, the effect on the extraction of $\ge$
decreases very slowly for large $Q^2$ values. While Coulomb distortion
explains only a portion of the discrepancy and appears to be small compared
to the effect of two hard photon exchange~\cite{blunden03, chen04}, its
inclusion is rather straightforward and should be done.

\begin{acknowledgments}

This work was supported by the U. S. Department of Energy, Office of
Nuclear Physics, under contract W-31-109-ENG-38, and the Schweizerische
Nationalfonds.

\end{acknowledgments}
\bibliography{pcc}

\end{document}